\documentclass[superscriptaddress,twocolumn, showpacs,preprintnumbers,amsmath,amssymb,aps,nofootinbib]{revtex4}
\pdfoutput=1

\usepackage{amsmath,amssymb,amsthm,amsfonts}
\usepackage{mathrsfs}
\usepackage{slashed}
\usepackage{graphicx}
\usepackage{subfigure}
\usepackage{color,rotating}
\usepackage{setspace}
\usepackage{verbatim}
\usepackage{hyperref}

\graphicspath{{./figures/}}

\newcommand{\mywidth}{1} 
\newcommand{\AL}{\mathscr{A}}
\newcommand{\AT}{A}
\newcommand{\beq}{\begin{equation}}
\newcommand{\eeq}{\end{equation}}
\newcommand{\beqa}{\begin{eqnarray}}
\newcommand{\eeqa}{\end{eqnarray}}
\newcommand{\beqw}{\begin{widetext}\begin{equation}}
\newcommand{\eeqw}{\end{equation}\end{widetext}}
\newcommand{\beqaw}{\begin{widetext}\begin{eqnarray}}
\newcommand{\eeqaw}{\end{eqnarray}\end{widetext}}
\newcommand{\bfc}{\begin{figure}[h!]\begin{center}}
\newcommand{\efc}{\end{center}\end{figure}}

\newcommand{\tn}{\textnormal}
\newcommand{\mn}{\mu \nu}
\newcommand{\mnr}{\mu \nu \rho}

\def\0#1#2{\frac{#1}{#2}}
\def\eq#1{eq.~(\ref{#1})}

\def\fig#1{FIG.~\ref{#1}}
\def\Fig#1{FIG.~\ref{#1}}

\def\CA{{\mathcal A}}
\def\CB{{\mathcal B}}
\def\CC{{\mathcal C}}
\def\CD{{\mathcal D}}
\def\CE{{\mathcal E}}
\def\CF{{\mathcal F}}

\begin{document}
\title{Exploratory study of the temperature dependence of magnetic vertices in $SU(2)$ Landau gauge Yang--Mills theory}
\author{Leonard Fister} 
\affiliation{Institut de Physique Th\'eorique, CEA Saclay, F-91191 Gif-sur-Yvette, France}
\author{Axel Maas}
\affiliation{Institute for Theoretical Physics, University of Jena, Max-Wien-Platz 1, D-07743 Jena, Germany}
\affiliation{Institute of Physics, University of Graz, Universit\"atsplatz 5, A-8010 Graz, Austria (present address)}

\begin{abstract}
Vertices describe the interactions between the fundamental degrees of freedom, and are therefore of vital importance in many \textit{ab initio} descriptions of field theory, especially using functional methods. To this end, we present the first lattice study of the thermal behavior of (minimal) Landau-gauge $SU(2)$ Yang--Mills three-point functions, i.e.\ three-gluon and ghost-gluon vertices. Focusing on the chromomagnetic sector, we find that the phase transition mainly affects the three-gluon vertex, while the ghost-gluon vertex is relatively inert.
\end{abstract}


\maketitle

\section{Introduction}\label{sec:introduction}

The study of heavy-ion collisions at RHIC and CERN have demonstrated that QCD is substantially more complex at and for some temperature range around the crossover than originally anticipated; see, e.g., \cite{Ullrich:2013qwa}. While lattice calculations have unraveled much of the static properties of the produced strongly interacting matter \cite{Bhattacharya:2014ara,Cheng:2009be,Borsanyi:2013bia}, many of the dynamical properties are still not completely understood. These questions include, e.g., the fast thermalization or the collective behavior of soft pions \cite{Ullrich:2013qwa}. Thus, ultimately an \textit{ab initio} nonequilibrium description will be necessary. This is, however, a challenging problem. One possible approach, for which real-time behavior becomes more and more accessible, are functional methods \cite{Strauss:2012dg,Wambach:2014vta,Haas:2013hpa,Berges:2012ty}. Still, these methods require approximations. To check and improve these approximations already static information is valuable. To obtain 
such information is the aim of the present paper. 

This will be achieved by the study of the chromomagnetic three-point vertices. To understand their importance, recall that all information of any quantum field theory is encoded in its correlation functions. Since vertices describe the basic interactions between the fundamental degrees of freedom, they provide access to conceptual questions, bound states, and even phenomenological observables  \cite{Alkofer:2000wg,Fischer:2008uz, Binosi:2009qm, Maas:2011se, Boucaud:2011ug, Vandersickel:2012tz}. In this work, we study, for the first time, and thus on an exploratory level, the temperature dependence of the three-point functions of $SU(2)$ Yang--Mills theory in (minimal) Landau gauge, i.e.\ the three-gluon and ghost-gluon vertices, for a range of different kinematic configurations. The three-point functions are of particular importance, since in most cases the approximations in functional calculations rely on the ordering of correlation functions in the sense that only $n$-point functions up to a certain order are included. 
Until recently, even in the vacuum only the gluon and ghost propagators could be resolved fully, whereas approximations for $n$-point functions with $n\geq 3$ have been made \cite{Alkofer:2000wg,Fischer:2008uz,Binosi:2009qm, Maas:2011se, Boucaud:2011ug, Vandersickel:2012tz}.

Already when including three-point vertices self-consistently, the level of complexity encountered in functional equations becomes much higher. As a consequence, vertex computations have been so far mainly done at vanishing temperature \cite{Fister:2011uw, Fister:Diss, Huber:2012kd, Aguilar:2013xqa, Pelaez:2013cpa, Aguilar:2013vaa, Blum:2014gna, Eichmann:2014xya}, except for an incipient study on the thermal behavior of the ghost-gluon vertex \cite{Fister:2011uw, Fister:Diss}. The thermal behavior of the three-point functions presented here thus serves as an important step in finding appropriate truncations or, in addition, may even provide a guideline in the computation of vertices self-consistently in functional frameworks.

In addition, the thermal dependence of the vertices may help to resolve a pressing question concerning the deviation of finite temperature continuum results \cite{Fister:2011uw, Fister:2011um, Fister:Diss, Huber:2013yqa, Reinosa:2013twa} from lattice simulations \cite{Cucchieri:2007ta, Fischer:2010fx, Bornyakov:2010nc, Aouane:2011fv, Maas:2011ez, Cucchieri:2012nx, Silva:2013maa, Mendes:2014gva}: the (almost) independence of the ghost propagator with respect to temperatures from zero to well above the critical temperature $T_c\approx 277\,\tn{MeV}$ has not be recovered in functional computations yet. This may be due to an insufficient truncation of the temperature dependence of the three-point functions. As will be seen, the three-gluon vertex shows a dependence on the temperature  while the ghost-gluon vertex is almost independent of temperature. These behaviors, if confirmed in a more systematic investigation, may provide a clue for the aforementioned insensitivity of the ghost propagator.

The main aim here is to get a first idea of interesting regions to set the stage for more detailed investigations. Hence, this study is limited to rather coarse and small lattices. Spatial volumes are $N_x^3=20^3,\,30^3$, whereas the temporal extent around the phase transition region is $N_t=4$. For the propagators there are significant lattice artifacts at these settings \cite{Bornyakov:2010nc, Aouane:2011fv, Maas:2011ez, Cucchieri:2012nx, Silva:2013maa, Mendes:2014gva}, so we do not expect yet quantitatively fully reliable results. However, for vertices so far the qualitative features in the vacuum have been yet less sensitive to lattice artifacts than for the propagators \cite{Cucchieri:2006tf,Cucchieri:2008qm,Maas:2007uv,Ilgenfritz:2006he}.

This manuscript is structured as follows. In Sec.~\ref{sec:Vertices} we set the technical stage, supported by some details in Appendix~\ref{app:Simulationalsetup}. The results are presented in the main part of the text in Sec.~\ref{sec:Results}. Some final conclusions are drawn in Sec.~\ref{sec:Conclusions}. A limited study of the volume dependence of the results can be found in Appendix \ref{app:artifacts}. Since our volumes are much smaller than those employed for investigations of the propagators, we do not include results on the latter, but rather refer to the literature \cite{Cucchieri:2007ta, Fischer:2010fx, Bornyakov:2010nc, Maas:2011ez, Cucchieri:2012nx, Silva:2013maa, Mendes:2014gva}.

\section{Vertices}\label{sec:Vertices}

In this section we briefly recall computations of three-point vertices using lattice gauge theory. In particular, we highlight the differences between computations at zero and nonzero temperature. Our approach is a straightforward extension of the analysis at zero temperature of Refs.~\cite{Cucchieri:2006tf,Cucchieri:2008qm,Cucchieri:2004sq}, and therefore many of the details skipped here can be found in these references. The introduction of temperature, especially determining the actual temperature, is done in the same way as in Ref.~\cite{Maas:2011ez}. See also Appendix \ref{app:Simulationalsetup} for further technical details.

Correlation functions are gauge-dependent quantities. Thus, one has to define a gauge condition. Here, we will use the standard minimal Landau gauge\footnote{We do not intend our calculation to be an approximation to the results in the absolute Landau gauge, i.e., of a gauge fixing to the fundamental modular region. Hence, there is no Gribov noise, since Gribov copies are fully taken into account in the definition of the minimal Landau gauge \cite{Maas:2011se}. This should be kept in mind when comparing to results in gauges that treat Gribov copies differently.} \cite{Maas:2011se}, to compare directly with results at vanishing temperature. As a consequence, the gluon propagator, $D_{\mn}^{ab}(p)$, is four-dimensionally transverse. Since higher $n$-point functions are connected via gluon or ghost propagators, the latter ones being Lorentz-scalar particles, transverse structures close among themselves. As a consequence, only the transverse parts of any $n$-point function contributes to the dynamics, if no 
extreme singularities are encountered\cite{Fischer:2008uz}. In particular, on the lattice it is only possible to determine the transverse part of correlation functions \cite{Maas:2011se}. In addition, the Landau gauge is covariant, and thus O(4) invariance allows at zero temperature for dependence on the modulus of the 4-momentum only.

At nonzero temperature, however, the situation is more complicated. The heat bath, which we choose to be in temporal direction, permits distinguishing the directions orthogonal and parallel to it. This generates two structural changes.

First, Lorentz/O(4) symmetry is no longer manifest. As a consequence, the gluon propagator has to be spanned via two different tensor structures, the so-called (chromo)electric and (chromo)magnetic propagators, $G_{\AL,\mn}^{ab}(p)$ and $G_{\AT,\mn}^{ab}(p)$, respectively, in the three-dimensional spatial subspace. In contrast, the trivial tensor structure of the ghost propagator, $G^{ab}_c(p)$, is unaffected as it is a Lorentz scalar. Because the minimal Landau gauge does not constrain the global gauge degree of freedom, these propagators are color diagonal, which has been confirmed explicitly by several calculations \cite{Maas:2011se}. This also implies that for the gauge group $SU(2)$ the three-point vertices' color structure is completely determined by the structure constants.  

Second, the presence of the heat bath triggers separate dependence on temporal and spatial components of the 4-momenta: therefore, the dependence on a general 4-momentum, $p$, is in fact a separate dependence on spatial momentum, $\vec{p}$, and Matsubara frequencies, $p_0$, which are discrete due to the compactification in the (imaginary) time direction. For both gluons and ghosts the energies are given by $2\pi n T$, with $T$ the temperature, and $n$ integer. This distinguishes soft modes with $n=0$ and hard modes with $n\neq 0$. The latter show a rather trivial behavior for the propagators \cite{Fischer:2010fx}, and become irrelevant for the dynamics in the high-temperature limit. Anticipating a similar behavior for the vertices, we will concentrate here exclusively on the soft modes.

Turning to vertices now, the bases of possible tensors are richer. Here, we will only consider those vertex components proportional to the tree-level vertex. In contrast to the vacuum, where this restriction singles out exactly one tensor, at nonzero temperature the basis is enlarged because either magnetic or electric gluons may be attached to the vertices. However, because these vertices are proportional to the momenta, restricting to soft modes automatically singles out the chromomagnetic, i.e.\ purely spatial, part of the vertices. These are identical to the three-gluon vertex and ghost-gluon vertex of the dimensionally reduced theory in the infinite temperature limit \cite{Maas:2011se}. Thus, by construction, we have to approach the known results for the three-dimensional vertices \cite{Cucchieri:2006tf,Cucchieri:2008qm} in this limit. In fact, already at roughly $2T_c$, the results are not too far away from the infinite-temperature limit. This rapid approach is similar to what is found for the 
propagators.

Thus, we are concerned with two three-point vertices, the three-gluon and ghost-gluon vertices,
\beq
\Gamma^{(3),\,abc}_{\AT^3,\,\mu\nu\rho}\left(p,q,r\right)\,,\quad \textnormal{and}\quad 
\Gamma^{(3),\,abc}_{c \bar c\AT, \, \mu}(p,q,r)\,,
\label{eq:vertices}
\eeq
where the subscript $\AT$ indicates that the external gluons are magnetic. In \eq{eq:vertices} momentum conservation ensures that these vertices depend on two external momenta only, from which three independent kinematic variables can be formed in the vacuum: e.g., the moduli of the two 4-momenta and the angle between them, respectively. At nonzero temperature, due to differentiation between spatial momenta and Matsubara frequencies, the vertices depend on five independent kinematic variables, two Matsubara frequencies and the moduli of and angle between spatial momenta. For the soft mode, therefore, only three parameters derived from the spatial momenta remain. In the following, we study momentum configurations that are directly compatible with the hypercubic symmetry of the spatial lattice. As detailed in Refs.~\cite{Cucchieri:2006tf,Cucchieri:2008qm}, due to their importance in continuum approaches, we choose the \textit{orthogonal} configuration, with two momenta being  equal in magnitude and the angle 
between them being 90$^{\circ}$; the \textit{symmetric} one, with all external momenta equal with an angle of 60$^{\circ}$ between each pair of momenta; and a third configuration in which one (gluon) momentum vanishes exactly and the other two momenta are therefore at 180$^{\circ}$. Note that on a finite lattice it is not possible to study zero ghost momentum. Furthermore, because of the ghost-antighost symmetry in the Landau gauge \cite{Alkofer:2000wg}, it is sufficient to study the dependence on the ghost momentum. Furthermore, the Bose symmetry of the three-gluon vertex constrains its momentum dependence.

An important technical constrain is that, contrary to continuum approaches, on the lattice only full correlation functions are accessible but not vertex functions directly. The latter ones are parametrized by their dressing functions, $G^{\AT^3}(p,q)$ for the three-gluon vertex and $G^{c\bar c\AT }(p,q)$ for the ghost-gluon vertex, respectively. Schematically, we have\footnote{Note that there is only one tensor structure in the ghost-gluon vertex at zero temperature, and thus only one soft structure at finite temperature.}
\beqa
\Gamma^{(3),\,abc}_{\AT^3, \, \mnr}(p,q,r) &=& G^{\AT^3}(p,q) \Gamma^{\textnormal{tl},\AT^3,\, abc}_{\mnr}(p,q,r)\nonumber\\
&&+\text{other tensor structures} \nonumber\,,\\
\Gamma^{(3),\,abc}_{c\bar c\AT, \, \mu}(p,q,r) &=& G^{c\bar c\AT}(p,q) \Gamma^{\textnormal{tl},c\bar c\AT,\, abc}_{\mu}(p,q,r)\nonumber
\eeqa
where the $\Gamma^{\textnormal{tl},\AT^3,\, abc}_{\mnr}$ and $\Gamma^{\textnormal{tl},c\bar c\AT,\, abc}_{\mu}$ are the (color antisymmetric) classical tensor structures, including lattice artifacts \cite{Cucchieri:2006tf,Rothe:1992nt}. Access to the dressing functions is gained by amputation and projection via \cite{Cucchieri:2006tf}
\beq
G^{\phi^3} = 
\frac{\Gamma^{(3)}_{\CA\CB\CC} G_{\CA\CD}G_{\CB\CE}G_{\CC\CF} \Gamma^{\textnormal{tl}}_{\CD\CE\CF}}
{\Gamma^{\textnormal{tl}}_{\CA\CB\CC} G_{\CA\CD}G_{\CB\CE}G_{\CC\CF} \Gamma^{\textnormal{tl}}_{\CD\CE\CF}}\,,
\label{eq:projection}
\eeq
where $G^{\phi^3}$ can be either $G^{\AT^3}$ or $G^{c\bar c\AT}$, the calligraphic letters are multi-indices comprising field-type, Lorentz and color indices as well as spatial momenta. We stress here that in \eq{eq:projection} no summation over Matsubara frequencies is involved because we only consider the zero mode, hence, higher modes drop out. Note that both dressing functions defined in \eq{eq:projection} are dimensionless.

The propagators are calculated in the same way as in previous studies at finite temperature \cite{Cucchieri:2007ta} and are, of course, along the corresponding spatial momenta. Thus, the longitudinal dressing function does not contribute to the normalization of the magnetic vertices.

Note that we keep the number of points in the time direction around the phase transition fixed and just vary $\beta$ and thus the lattice spacing. Therefore, the spatial volumes shrink slightly while increasing the temperature. Since the propagators are volume dependent (as to some extent most likely the vertices too), some volume dependence is intermingled with the temperature dependence. However, the observed effects are substantially  larger than the volume dependence of the vertices at zero or infinite temperature, so we are rather confident that we also see genuine temperature effects.

Finally, in principle there is a (finite) renormalization possible. At the present lattice discretization, these renormalization constants are found to be only slightly dependent on $\beta$. Thus, we do not perform any renormalization here, except below in Fig.~\ref{fig:g3v-1m}.

\section{Results}\label{sec:Results}

In this section we present results on the thermal behavior of the magnetic three-point functions in comparison to their behavior at vanishing \cite{Cucchieri:2008qm} and infinite\cite{Cucchieri:2006tf, Cucchieri:2008qm} temperature. The behavior of the vertices is qualitatively similar on both spatial lattice volumes $N_x^3=20^3$ and $N_x^3=30^3$ studied here. Hence, we restrict the discussion in the main text to the larger lattice volume and defer the study of the volume dependence to Appendix~\ref{app:artifacts}. We will study the two vertices separately in the following two subsections. We have not investigated discretization effects, as this would require finer lattices, and thus substantially more computing time to keep the same spatial volumes. We note, however, that such effects have a substantial impact on the propagators, and should be addressed in any next step beyond this exploratory one  \cite{Bornyakov:2010nc, Maas:2011ez, Cucchieri:2012nx, Silva:2013maa, Mendes:2014gva}.

\subsection{Ghost-gluon vertex}

In the vacuum the $SU(2)$ ghost-gluon vertex remains almost constant with deviations of the order of 30\% from its tree-level part, in agreement between functional methods and lattice calculations \cite{Schleifenbaum:2004id, Fister:2011uw, Fister:Diss, Huber:2012kd, Aguilar:2013xqa, Pelaez:2013cpa, Aguilar:2013vaa,Cucchieri:2008qm}.

\begin{figure*}\centering
\includegraphics[width=\mywidth\textwidth]{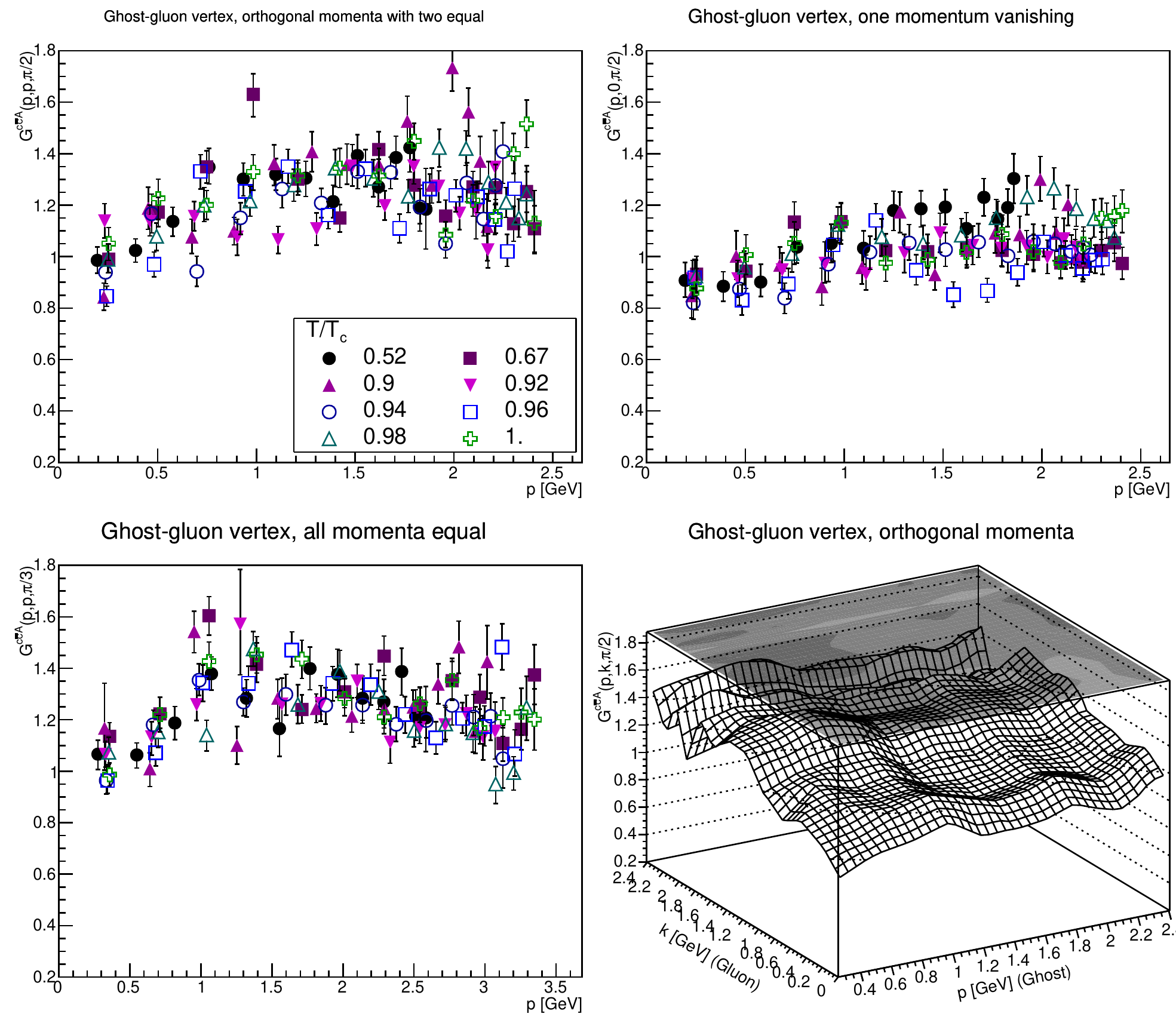}
\caption{Various kinematic configurations of the ghost-gluon vertex at temperatures below criticality.  The lower right graph illustrates the vertex at the critical temperature as a function of ghost and gluon momentum.}
\label{fig:ggv_lowT}
\end{figure*}

The thermal dependence of the ghost-gluon vertex for temperatures $T\leq T_c$, is shown in \fig{fig:ggv_lowT}. Within the statistical uncertainties, no pronounced temperature dependence is observed, and the vertex remains essentially tree level for the temperature range studied.

\begin{figure*}\centering
\includegraphics[width=\mywidth\textwidth]{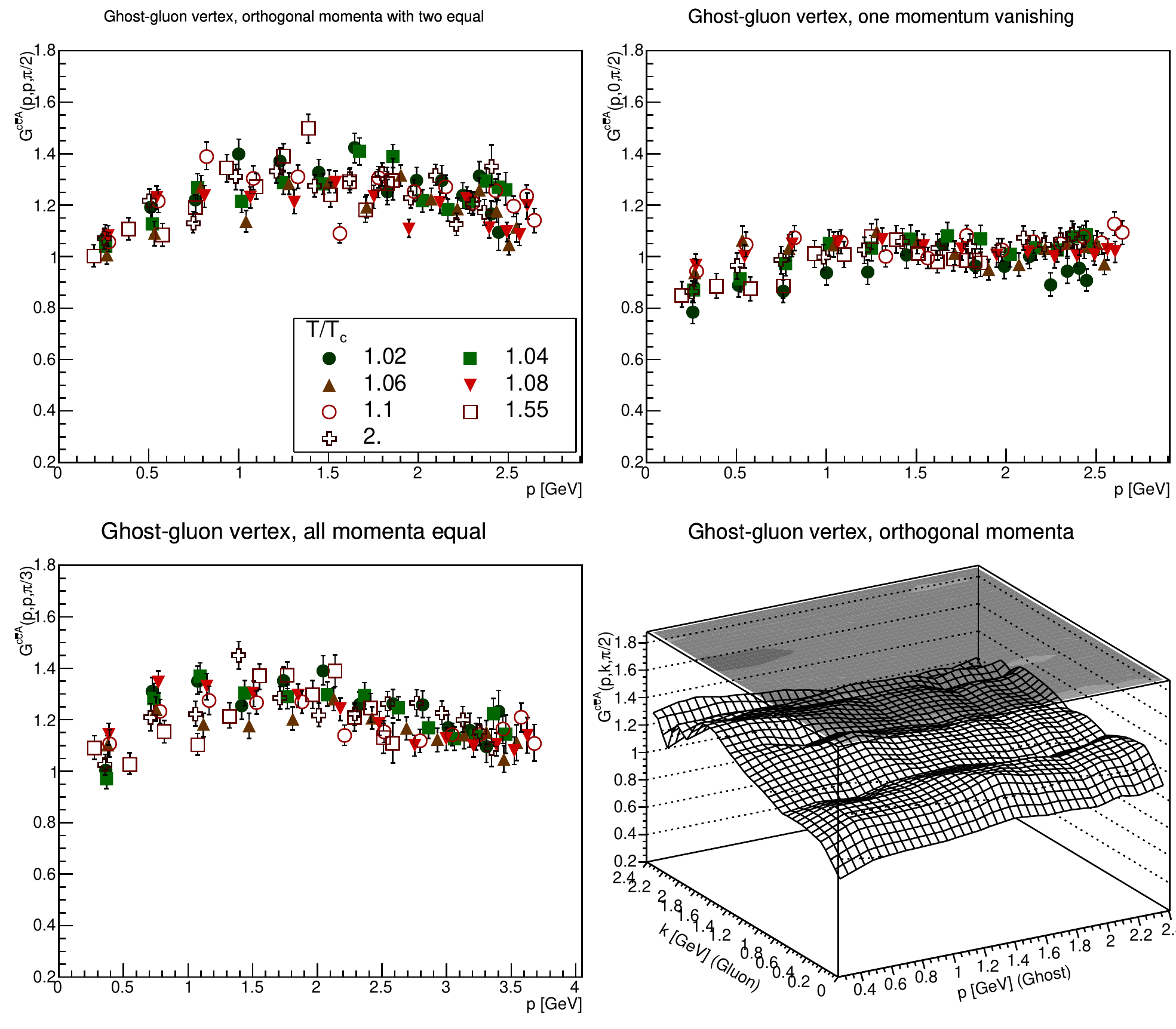}
\caption{Ghost-gluon vertex as in Fig.~\ref{fig:ggv_lowT}, however, for temperatures above the transition. The lower right plot is at the largest temperature.}
\label{fig:ggv_highT}
\end{figure*}

For temperatures above the phase transition, $T> T_c$, the results are shown in \fig{fig:ggv_highT}. Again, within statistical accuracy almost no temperature dependence is observed. In this way, the ghost-gluon vertex shows a behavior quite similar to the soft mode of the ghost propagator itself \cite{Maas:2011ez}, which is also essentially independent of temperature.

\begin{figure*}\centering
\includegraphics[width=\mywidth\textwidth]{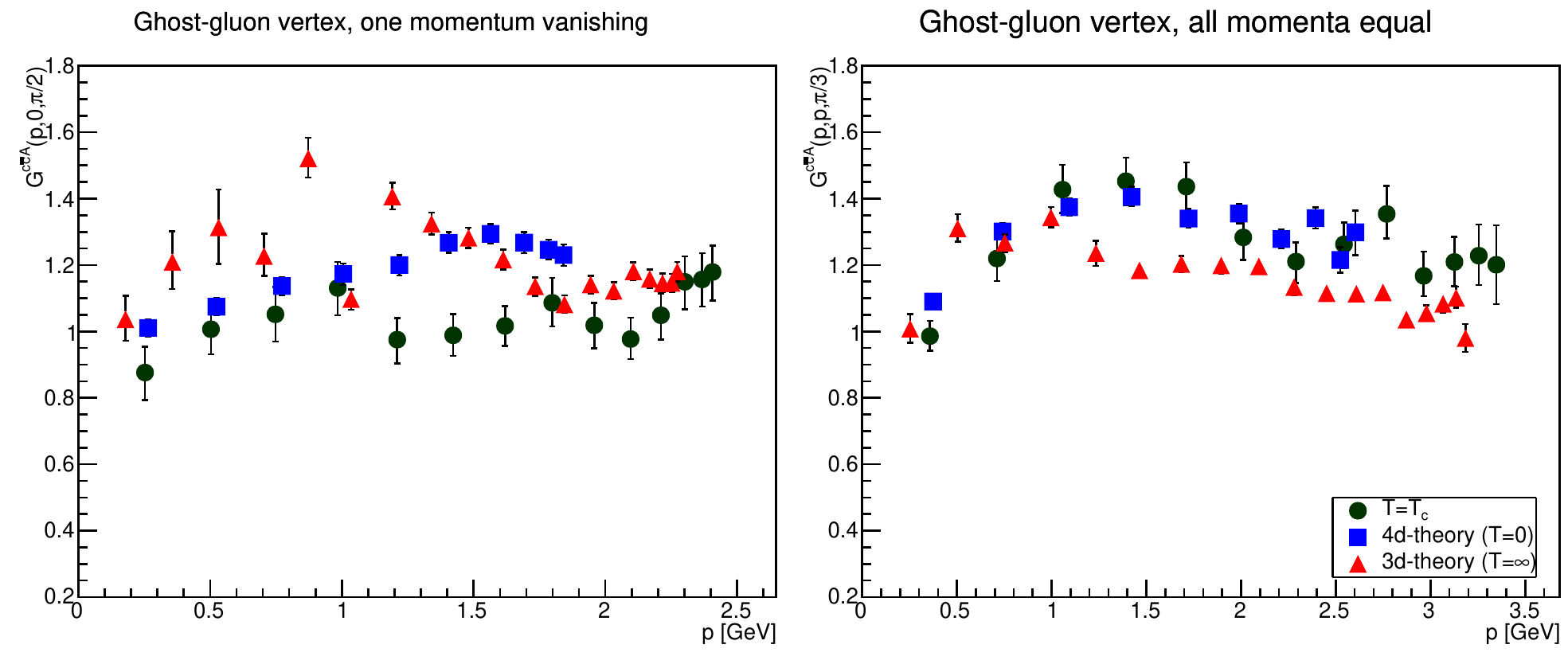}
\caption{The ghost-gluon vertex at the phase transition compared to the results from zero and infinite temperature from Ref.~\cite{Cucchieri:2008qm}.}
\label{fig:ggv-30_limit_3d}
\end{figure*}

At very high temperature dimensional reduction forces the vertex to approach its three-dimensional limit. However, because of the almost temperature independence of the vertex, this is essentially true for all temperatures. This is emphasized in \Fig{fig:ggv-30_limit_3d}, where the vertex at the critical temperature is compared to the zero and infinite-temperature limit from Ref.~\cite{Cucchieri:2008qm}. As is visible, within the comparatively large statistical uncertainties, the vertex is essentially the same. Note that for the comparison to the three-dimensional theory, the string tension in four and three dimensions has been set to the same dimensionful quantity \cite{Cucchieri:2008qm}.

\subsection{Three-gluon vertex}

At vanishing temperature the three-gluon vertex is significantly suppressed at intermediate and low momenta compared to its tree-level value in both four and three dimensions \cite{Cucchieri:2006tf, Cucchieri:2008qm}. However, statistics and volumes are insufficient to specify yet the infrared behavior. Nevertheless, the data are compatible with both functional results, where the vertex turns negative in the deep infrared \cite{Fister:Diss, Blum:2014gna, Eichmann:2014xya}, as well as lower-dimensional results where in larger volumes the same behavior was first seen \cite{Cucchieri:2008qm,Maas:2007uv}.

\begin{figure*}\centering
\includegraphics[width=\mywidth\textwidth]{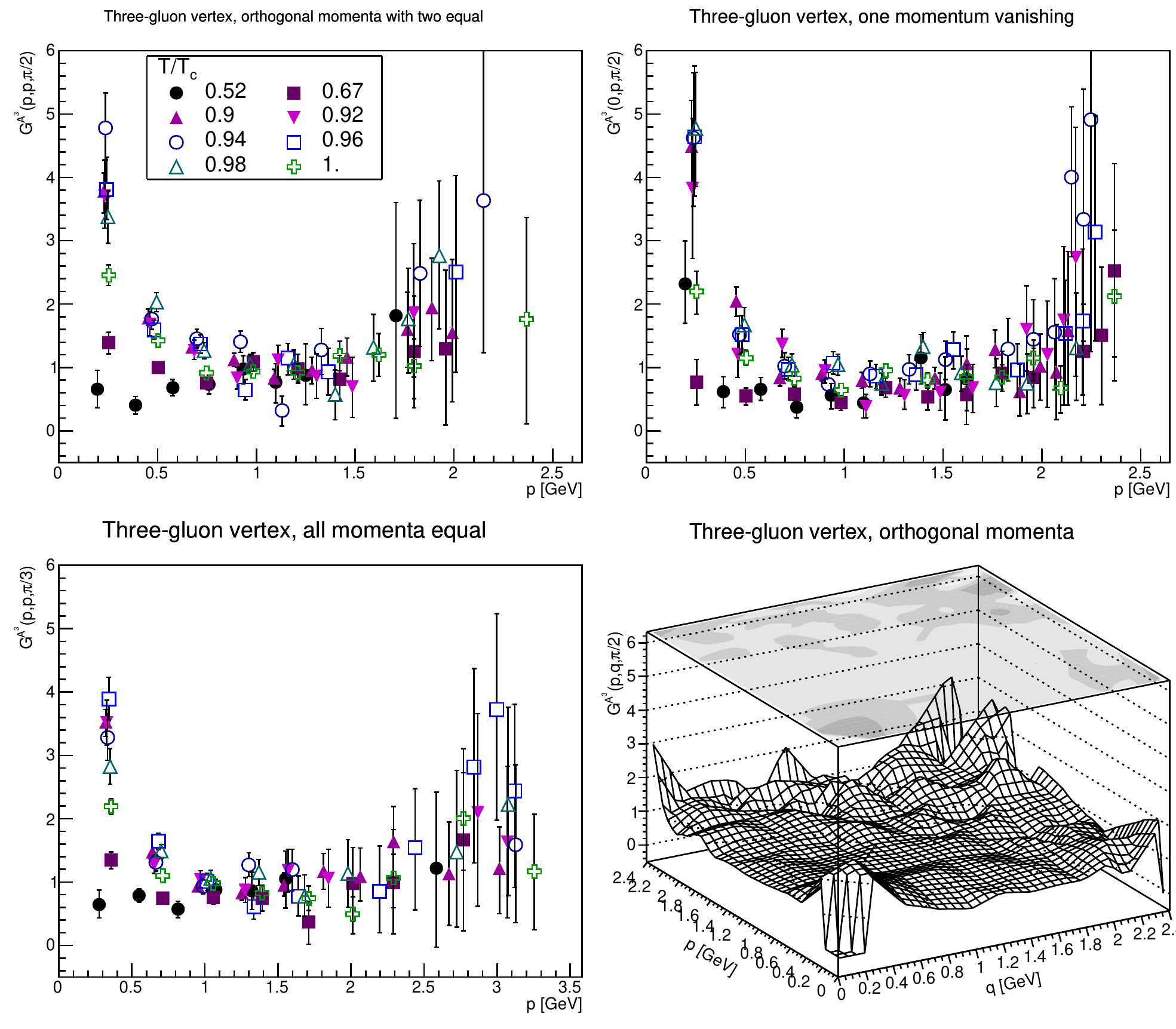}
\caption{Various kinematic configurations of the three-gluon vertex at temperatures below criticality. The lower right graph illustrates the vertex at the highest temperature in the figure as a function of two independent momenta. Only data with relative error smaller than 1 are shown.}
\label{fig:g3v_lowT}
\end{figure*}

The strong statistical fluctuations are also the bane of the present calculations, especially at high momenta. However, a very interesting observation can be made at low momenta below the phase transition, see \Fig{fig:g3v_lowT}. Here, the vertex becomes substantially enhanced toward the infrared when approaching the phase transition. However, as in the case of the longitudinal propagator \cite{Maas:2011ez,Mendes:2014gva}, the maximum enhancement occurs not at the phase transition, but slightly below, around 0.95$T_c$.

\begin{figure*}\centering
\includegraphics[width=\mywidth\textwidth]{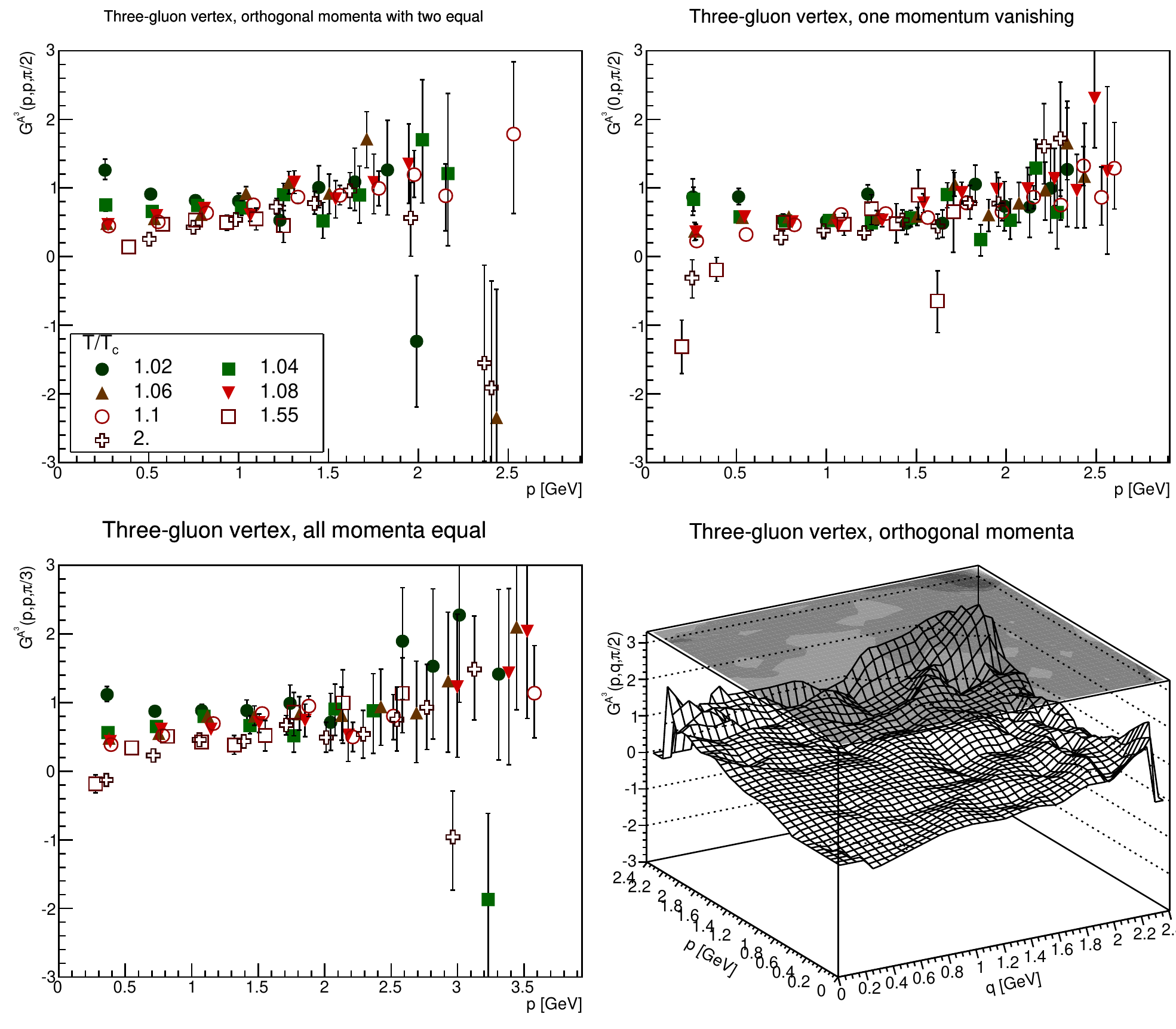}
\caption{Three-gluon vertex as in \Fig{fig:g3v_lowT}, however, for temperatures above criticality.}
\label{fig:g3v_highT}
\end{figure*}

\begin{figure*}\centering
\includegraphics[width=\mywidth\textwidth]{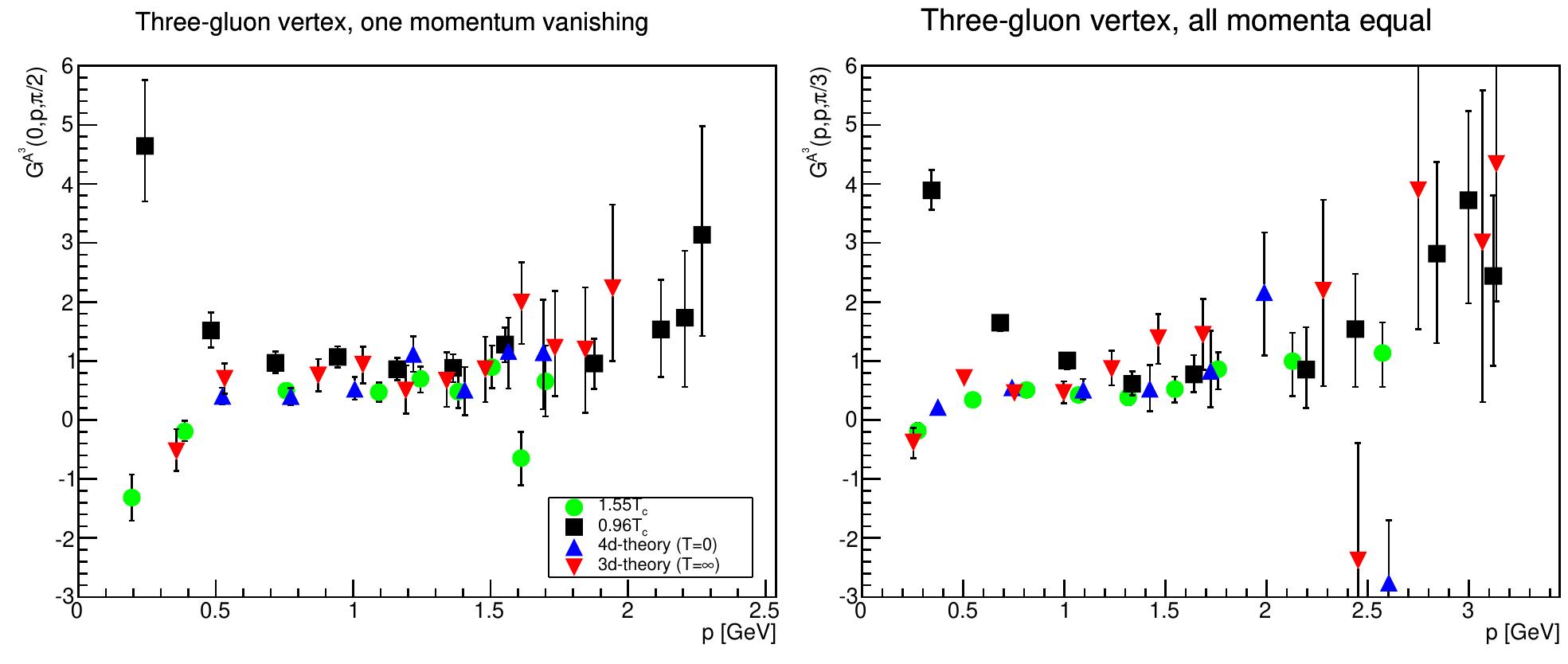}
\caption{The three-gluon vertex at the point of maximum enhancement at $0.96T_c$ and at 1.55$T_c$, compared to the results from zero and infinite temperature from Ref.~\cite{Cucchieri:2008qm}. Only data with relative error smaller than 1 are shown.}
\label{fig:g3v-30_limit_3d}
\end{figure*}

This suppression is essentially immediately gone in the high-temperature phase, as is seen in \Fig{fig:g3v_highT}. In fact, already at about $1.5T_c$, the original suppression is there, and even a sign change can be observed. Thus, the behavior in the infinite-temperature-limit is recovered \cite{Cucchieri:2008qm}, as shown in \Fig{fig:g3v-30_limit_3d}. This is, of course, a dramatic dependence on the temperature, inducing a qualitative change. Especially that the vertex stays positive is a genuine feature of the three-point correlation function, since the required normalization by the propagators in \eq{eq:projection} cannot alter the sign, just the magnitude.

\begin{figure*}\centering
\includegraphics[width=\mywidth\textwidth]{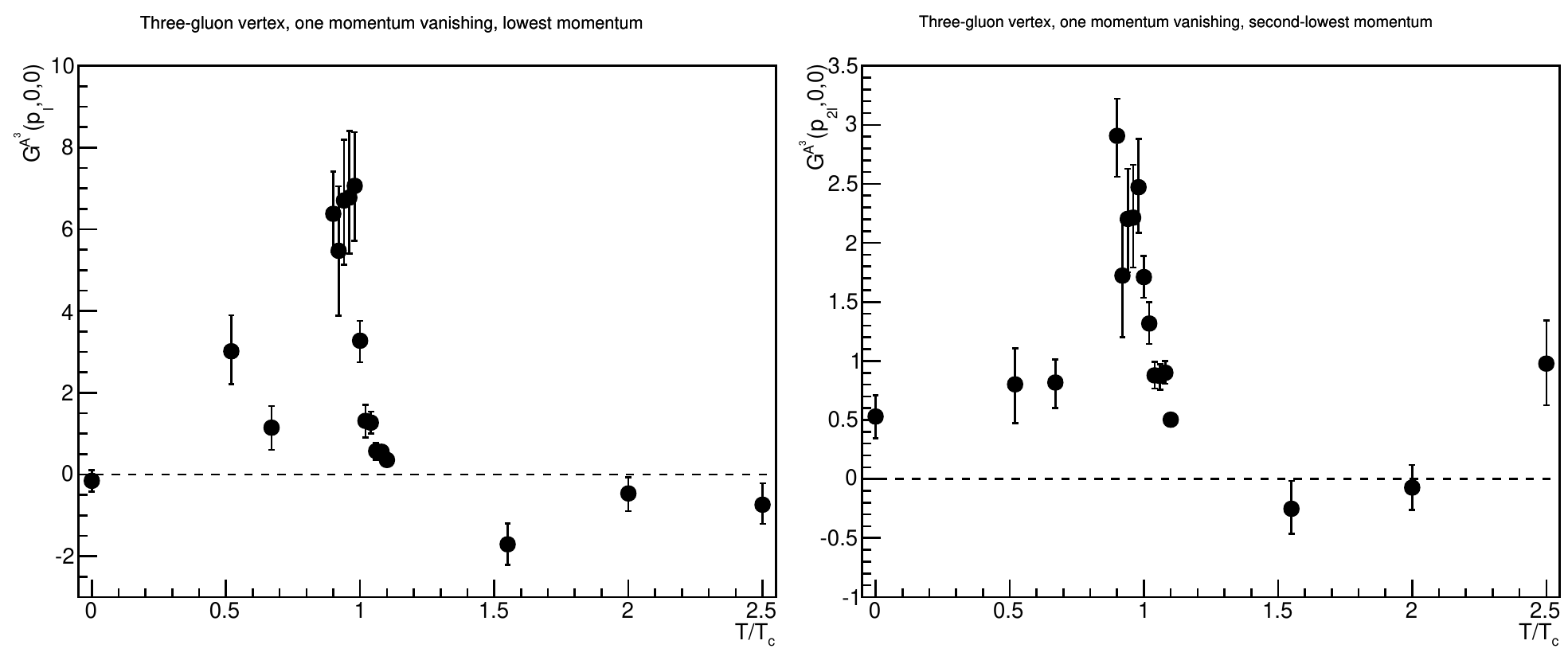}
\caption{The three-gluon vertex at the lowest momentum (left panel, 235$\pm$35 MeV) and next-to-lowest momentum (right panel, 370$\pm$80 MeV) as a function of temperature for the case of one gluon momentum vanishing. The variation in momenta stems from the different $\beta$ values. The vertex has been renormalized at 1.3 GeV to 1, using a linear extrapolation in $\beta$ of the zero-temperature data from Ref.~\cite{Cucchieri:2008qm}. The infinite-temperature results are plotted for comparison at 2.5$T_c$. However, given the quite different scales and renormalization properties, any comparison should be made with care.}
\label{fig:g3v-1m}
\end{figure*}

\begin{figure*}\centering
\includegraphics[width=\mywidth\textwidth]{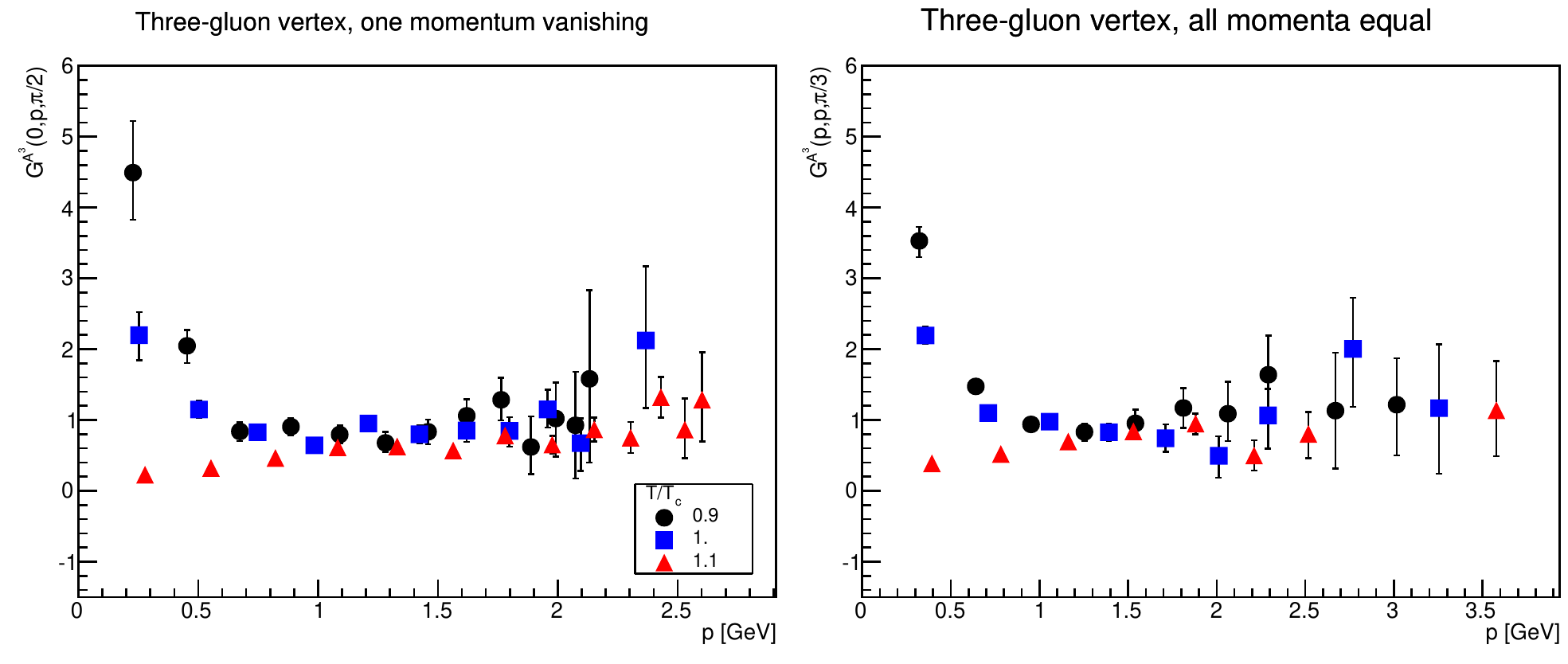}
\caption{The three-gluon vertex at temperatures around the phase transition for two momentum configurations. The infrared enhancement of the vertex observed for $T<T_c$ quickly disappears in the high-temperature phase.}
\label{fig:g3v-30_fewT}
\end{figure*}

To emphasize the point, we plotted the vertex at the lowest and next-to-lowest momenta in \Fig{fig:g3v-1m} as a function of temperature. For this direct comparison, we have renormalized the vertex at 1.3 GeV for one momentum vanishing, using a linear extrapolation of the more precise vacuum data from Ref.~\cite{Cucchieri:2008qm}. It is visible how the vertex becomes positive at some temperature substantially below the phase transition and then shows a pronounced peak or plateau structure directly below and up to the critical temperature. The drastic effect around the phase transition is again highlighted in \Fig{fig:g3v-30_fewT}. This behavior coincides with the one observed for the electric propagator \cite{Cucchieri:2007ta, Fischer:2010fx, Bornyakov:2010nc, Maas:2011ez, Cucchieri:2012nx, Silva:2013maa, Mendes:2014gva}. Whether it shows also a maximum rate of change very close to the phase transition, as is suggested by Ref.~\cite{Maas:2011ez} for the electric propagator, cannot be decided given the available precision.

We point out that this is unexpected, as in the soft three-gluon vertex only magnetic gluon fields enter. The reason is that for the soft case in \eq{eq:projection} the tree-level tensor structure is proportional to the momentum, and thus all time components, i.e., electric ones, vanish. Thus, only magnetic ones contribute. Since the magnetic propagator is rather inert as a function of temperature \cite{Cucchieri:2007ta, Fischer:2010fx, Bornyakov:2010nc, Maas:2011ez, Cucchieri:2012nx, Silva:2013maa, Mendes:2014gva}, no drastic change in the vertex suggested itself.

\section{Conclusions}\label{sec:Conclusions}

Summarizing, we have presented the first lattice investigation of the temperature dependence of the (chromo)magnetic, soft three-point vertices of $SU(2)$ Yang--Mills theory in minimal Landau gauge.

We find that the ghost-gluon vertex shows essentially no temperature dependence, similar to the ghost propagator itself. Given that only the ghost fields as well as chromomagnetic gluons, neither of which show a strong temperature dependence \cite{Cucchieri:2007ta, Fischer:2010fx, Bornyakov:2010nc, Maas:2011ez, Cucchieri:2012nx, Silva:2013maa, Mendes:2014gva}, enter into the correlation function, this is not really surprising.

However, the tree-level tensor dressing function of the three-gluon vertex shows a pronounced temperature dependence, with a strong positive infrared enhancement around the phase transition temperature. This is in stark contrast to the situation at zero or infinite temperature, in which the same dressing is strongly suppressed and likely even becomes negative. A similar sensitivity has been observed for the electric gluon propagator  \cite{Cucchieri:2007ta, Fischer:2010fx, Bornyakov:2010nc, Maas:2011ez, Cucchieri:2012nx, Silva:2013maa, Mendes:2014gva}. However, in the soft case studied here, no longitudinal gluon fields enter, but only magnetic ones. Thus, this outcome is quite unexpected.

Note, however, that the lattice volumes are rather small, and the discretizations are rather coarse. Both facts have quite a large impact on the propagators \cite{Bornyakov:2010nc, Maas:2011ez, Cucchieri:2012nx, Silva:2013maa, Mendes:2014gva}, and therefore these results should be taken with caution.

Ignoring this caution for one paragraph, and assuming that these results will hold in more detailed studies, the implications are quite significant. The absence of changes in the ghost-gluon vertex closes nicely with observations for propagators: within systematic errors due to volume and discretization effects, they are compatible with temperature-insensitive magnetic and ghost propagators. This should indicate the right avenue to understand the situation in functional calculations \cite{Fister:2011uw}. However, the strong changes of the three-gluon vertex indicate a subtle interplay in the gluonic sector at and around the phase transition. In fact, the infrared enhancement and absence of the sign change in the magnetic three-gluon vertex around the phase transition is quite unexpected.

Returning to a more adequate cautionary tone, the results are interesting but surely require more careful study of lattice artifacts. On the other hand, the magnetic three-gluon vertex shows that the magnetic sector is perhaps not as unaffected by the dynamics of the phase transition as originally anticipated by the behavior of the propagators. It thus appears that further studies of the magnetic vertices would be very helpful, also for functional calculations. Of course, given the behavior of the electric propagator and the present results, this begs the question on the behavior of the electric and mixed vertices. This is certainly a demanding project but nevertheless a necessary step toward an understanding of the dynamics around the phase transition in terms of the gluonic degrees of freedom.

{\it Acknowledgments} --- We thank Jan M. Pawlowski for discussions. L.F. is supported by the European Research Council under the Advanced Investigator Grant No. ERC-AD-267258 and Science Foundation Ireland Grant No. 11-RFP.1-PHY3193. A.M. is supported by the DFG under Grants No. MA 3935/5-1 and No. MA 3935/8-1 (Heisenberg program) and thanks the NUI Maynooth for hospitality. This work has been supported by Agence Nationale de la Recherche Project No. 11-BS04-015-01. The ROOT framework \cite{Brun:1997pa} has been used in this project.\\

\appendix

\section{Simulation setup}\label{app:Simulationalsetup}

The results in this work have been obtained as a generalization of Refs.~\cite{Cucchieri:2006tf, Cucchieri:2008qm, Fischer:2010fx}. We use a standard Wilson action for $SU(2)$ Yang--Mills theory. The configurations were generated using hybrid over-relaxation (HOR) updates by alternating five over-relaxation sweeps with one heat-bath sweep, the latter ones being done via a mixed Creutz and Kennedy--Pendleton algorithm. Before the first gauge-fixing $N_{\textnormal{init}}$ HOR updates were discarded for thermalization, with $N_{\textnormal{init}} = 2\left(10\,N_x+300\right)$, and between two consecutive measurements $N_{\textnormal{init}}/10$, HOR updates were performed. Gauge fixing to the minimal Landau gauge has been performed with an adaptive stochastic over-relaxation algorithm using a global stopping criterion \cite{Cucchieri:2006tf}. We use asymmetric lattices $N_t\times N_x$ and vary $\beta$ to scan the temperature around the phase transition. To map to physical temperatures, we obtain the string tension as a 
function of $\beta$ by interpolating the results of Ref.~\cite{Fingberg:1992ju} and setting the string tension to $(440$ MeV$)^2$, to be compatible with previous works, especially Ref.~\cite{Maas:2011ez}. The particular lattice settings employed are listed in Table~\ref{tab:simuls}.

\begin{table*}[!ht] 
\begin{tabular}{| c | c | c | c | c | c || c | c | c | c | c | c | c | c | }
\hline
$T/T_c$ & $\beta$ & $N_t$ & $N_x$ & 
$N_{\textnormal{conf.}}^{\AT^3}$ ($N_{\textnormal{runs}}^{\AT^3}$)  & $N_{\textnormal{conf.}}^{c\bar c\AT}$ ($N_{\textnormal{runs}}^{c\bar c\AT}$)& $T/T_c$ & $\beta$ & $N_t$ & $N_x$ & 
$N_{\textnormal{conf.}}^{\AT^3}$ ($N_{\textnormal{runs}}^{\AT^3}$)  & $N_{\textnormal{conf.}}^{c\bar c\AT}$ ($N_{\textnormal{runs}}^{c\bar c\AT}$) \\
\hline
0.52 & 2.200 & 6 & 20 & 718 (120) & 228 (40) & 0.52 & 2.200 & 6 & 30 & 1176 (290) & 164 (40) \\
0.67 & 2.299 & 6 & 20 & 744 (120) & 252 (40) & 0.67 & 2.299 & 6 & 30 & 1152 (290) & 156 (40) \\
0.90 & 2.260 & 4 & 20 & 747 (120) & 244 (40) & 0.90 & 2.260 & 4 & 30 & 3251 (840) & 154 (40) \\
0.92 & 2.268 & 4 & 20 & 718 (120) & 235 (40) & 0.92 & 2.268 & 4 & 30 & 1141 (289) & 180 (40) \\
0.94 & 2.276 & 4 & 20 & 731 (120) & 226 (40) & 0.94 & 2.276 & 4 & 30 & 1151 (290) & 171 (40) \\
0.96 & 2.284 & 4 & 20 & 735 (120) & 240 (40) & 0.96 & 2.284 & 4 & 30 & 1140 (285) & 163 (40) \\
0.98 & 2.292 & 4 & 20 & 707 (120) & 244 (40) & 0.98 & 2.292 & 4 & 30 & 1142 (286) & 165 (40) \\
1.00 & 2.299 & 4 & 20 & 718 (120) & 248 (40) & 1.00 & 2.299 & 4 & 30 & 3348 (840) & 160 (40) \\
1.02 & 2.306 & 4 & 20 & 708 (120) & 217 (40) & 1.02 & 2.306 & 4 & 30 & 1585 (212) & 246 (40) \\
1.04 & 2.312 & 4 & 20 & 998 (120) & 345 (40) & 1.04 & 2.312 & 4 & 30 & 1952 (244) & 320 (40) \\
1.06 & 2.319 & 4 & 20 & 1283 (120) & 456 (40) & 1.06 & 2.319 & 4 & 30 & 1952 (244) & 320 (40) \\
1.08 & 2.326 & 4 & 20 & 1440 (120) & 480 (40) & 1.08 & 2.326 & 4 & 30 & 1952 (244) & 320 (40) \\
1.10 & 2.332 & 4 & 20 & 1440 (120) & 480 (40) & 1.10 & 2.332 & 4 & 30 & 6272 (840) & 320 (40) \\
1.55 & 2.200 & 2 & 20 & 1440 (120) & 480 (40) & 1.55 & 2.200 & 2 & 30 & 2200 (275) & 320 (40) \\
2.00 & 2.299 & 2 & 20 & 1440 (120) & 480 (40) & 2.00 & 2.299 & 2 & 30 & 2216 (277) & 320 (40) \\
\hline
\end{tabular}
\caption[]{The lattice setups employed. The temperature is given in units of the critical temperature $T/T_c$, with $T_c\approx 277\,\textnormal{MeV}$, the lattice coupling $\beta$ and the number of lattice sites $N_t$ ($N_x$) in the temporal (spatial) direction. The average of the three-gluon (ghost-gluon) vertex is taken over $N_{\textnormal{conf.}}^{\AT^3}$ ($N_{\textnormal{conf.}}^{c\bar c\AT}$)  configurations, which were generated in $N_{\textnormal{runs}}^{\AT^3}$ ($N_{\textnormal{runs}}^{c\bar c\AT}$) individual runs.}
\label{tab:simuls}
\end{table*}

\section{Volume artifacts}\label{app:artifacts}

\begin{figure*}\centering
\includegraphics[width=\mywidth\textwidth]{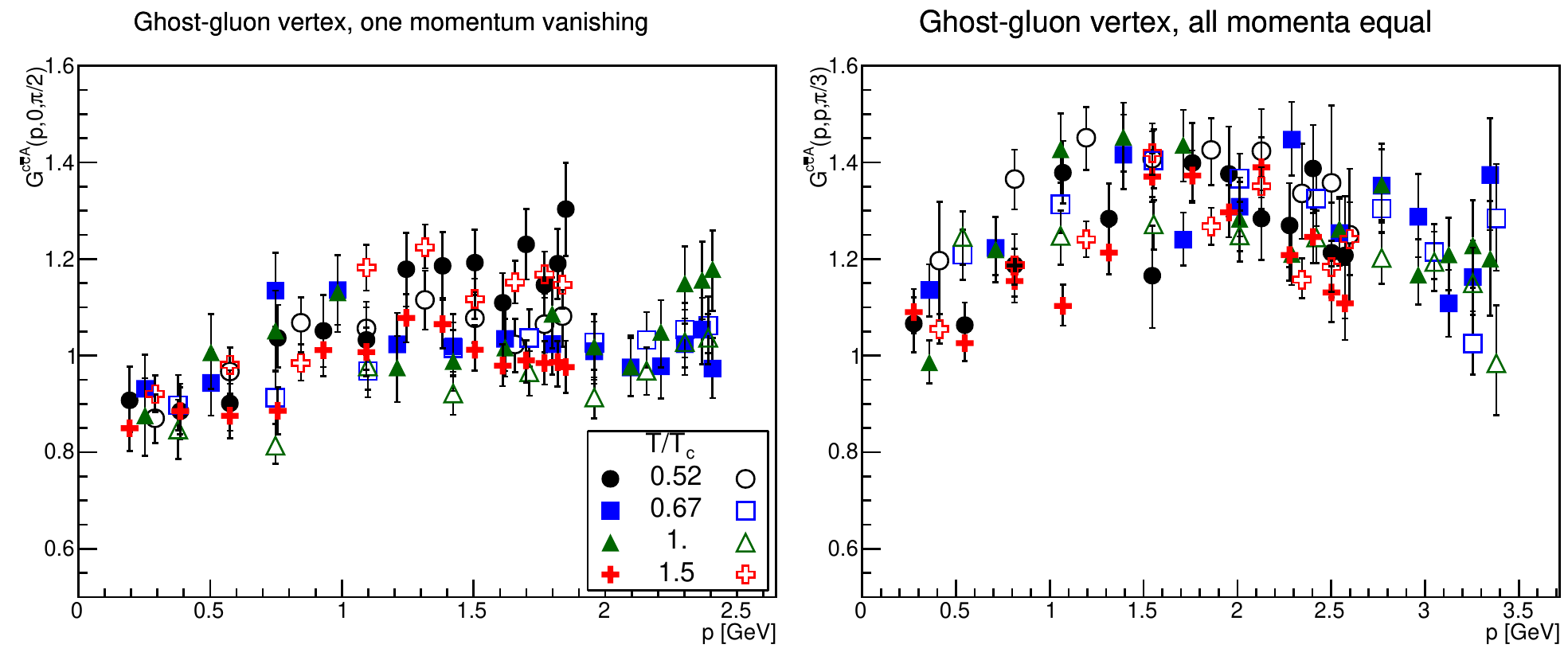}
\caption{Volume dependence of ghost-gluon vertex. Filled (empty) symbols denote the spatial lattice of size $N_x^3=30^3$ $(20^3)$.}
\label{fig:ggv-20-30}
\end{figure*}

Simulations were done on two different lattices with spatial volumes of $N_x^3=20^3,\,30^3$, see Table~\ref{tab:simuls}, but fixed lattice spacing. Thus, we can only study the volume dependence of our results but not yet the discretization dependence. The results for selected temperatures for the two volumes for the ghost-gluon vertex, shown in \fig{fig:ggv-20-30}, show essentially no dependence on the volume within the statistical error. The same applies to all other temperatures. Thus, within our systematic reach, the ghost-gluon vertex appears volume independent.

\begin{figure*}\centering
\includegraphics[width=\mywidth\textwidth]{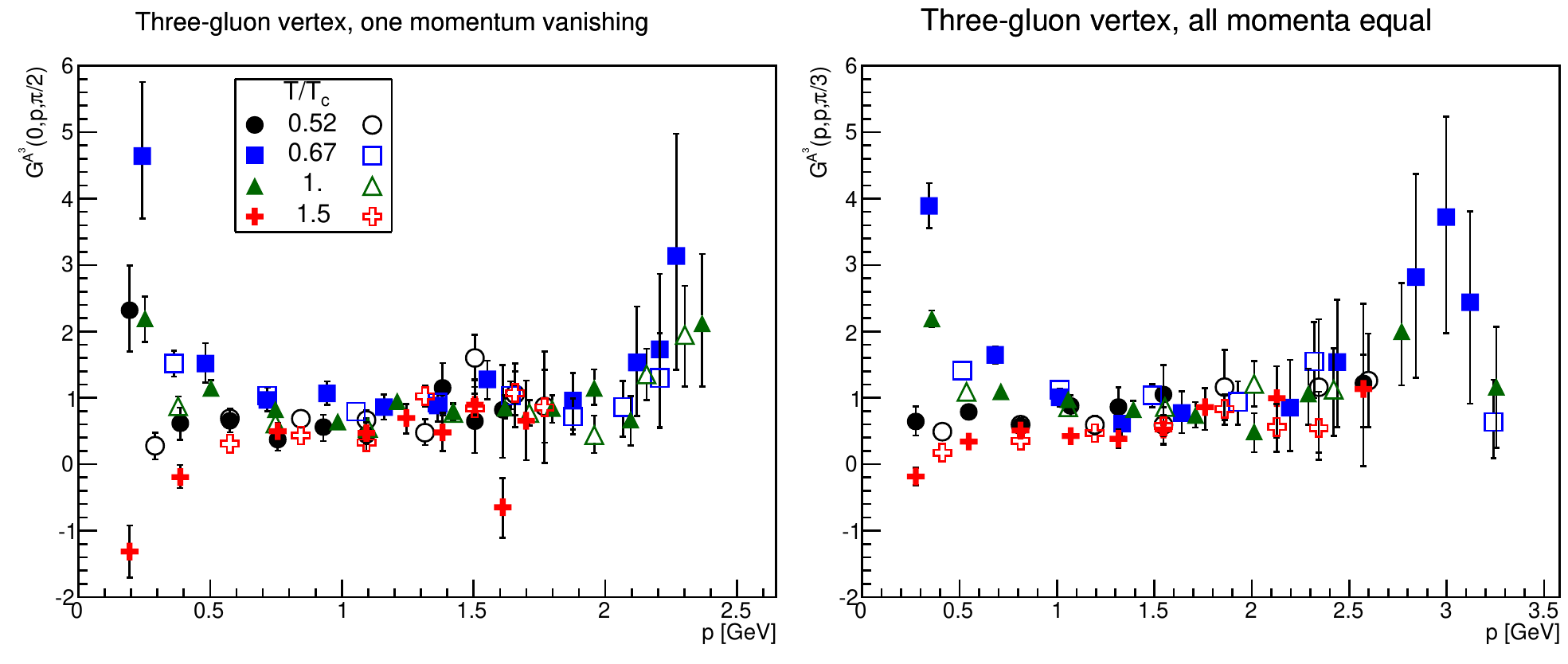}
\caption{Volume dependence of three-gluon vertex as in \Fig{fig:ggv-20-30}. Only data with relative error smaller than 1 are shown.}
\label{fig:g3v-20-30}
\end{figure*}

In case for the three-gluon vertex, shown in \Fig{fig:g3v-20-30}, we do see some volume dependence. However, increasing the volume further amplifies the enhancement observed in the main text, especially at the point of maximum enhancement. Also, at high temperatures, the suppression becomes stronger with increasing volume. Thus, the volume dependence favors the behavior described in the main text. However, here a word of caution is mandatory: For the gluon propagator at zero temperature, the volume dependence on volumes much larger than the present ones is actually qualitatively different than in the case of the present volumes (see Ref.~\cite{Maas:2011se} for a review). Hence, it cannot be excluded that also for the vertex a qualitative change may occur at much larger volumes. Given the amount of statistical fluctuations, this will require substantially more computing power than employed in the present calculation and is therefore left until such resources become available.

\bibliographystyle{./bibstyle}
\bibliography{./3PtFinT}

\end{document}